\documentclass[11pt]{article}

\usepackage{graphicx}

\begin{document}

\thispagestyle{empty}

\onecolumn

\begin{flushright}
{\large
SLAC--AP--139\\
February 2002\\}
\end{flushright}

\vspace{.8cm}

\begin{center}

{\LARGE\bf
Study of the NLC Linac Optics Compatible\\[2mm]
with a Low Energy Scenario~\footnote
{\normalsize
{Work supported by Department of Energy contract  DE--AC03--76SF00515.}}}

\vspace{1cm}

\large{
Y.~Nosochkov and T.O.~Raubenheimer\\
Stanford Linear Accelerator Center, Stanford University,
Stanford, CA 94309}

\end{center}

\vfill

\begin{center}
{\LARGE\bf
Abstract }
\end{center}

\begin{quote}
\large{
We explore the NLC linac optics compatible with a low energy scenario where
initially only part of the full linac is installed.  Optics modification
suitable for a low energy beam running and upgrade to the nominal energy is
discussed.  Linac parameters and beam tolerances in the modified lattice are
compared to the nominal design.
}
\end{quote}

\vfill

\newpage

\pagenumbering{arabic}

\pagestyle{plain}

\renewcommand{\thefootnote}{\fnsymbol{footnote}}

\begin{center}
{\bf\Large
Study of the NLC Linac Optics Compatible\\
with a Low Energy Scenario
\footnote{Work supported by Department of Energy contract 
DE--AC03--76SF00515.}}
\end{center}

\begin{center}
Y.~Nosochkov and T.O.~Raubenheimer\\
Stanford Linear Accelerator Center, Stanford University,
Stanford, CA 94309
\end{center}

\begin{center}
{\bf\large
Abstract }
\end{center}

\begin{quote} 

We explore the NLC linac optics compatible with a low energy scenario where
initially only part of the full linac is installed.  Optics modification
suitable for a low energy beam running and upgrade to the nominal energy is
discussed.  Linac parameters and beam tolerances in the modified lattice are
compared to the nominal design.

\end{quote}

\section{Introduction}

We consider a scenario where the NLC~\cite{nlc} is initially built for a lower
beam energy than in the nominal design.  The advantage is an earlier start of
machine operation at a lower cost since only part of the full linac may be
initially installed.  However, the linac design has to be compatible with the
low energy operation and upgrade to the nominal energy.  A modification of the
linac lattice suitable for the low energy option is discussed.  Parameters of
the modified lattice and ground motion tolerances are compared with the
nominal design (based on linac version May 2000).

\section{Optics}

\begin{figure}[tb]
\centering
\includegraphics*[width=80mm, angle=-90]{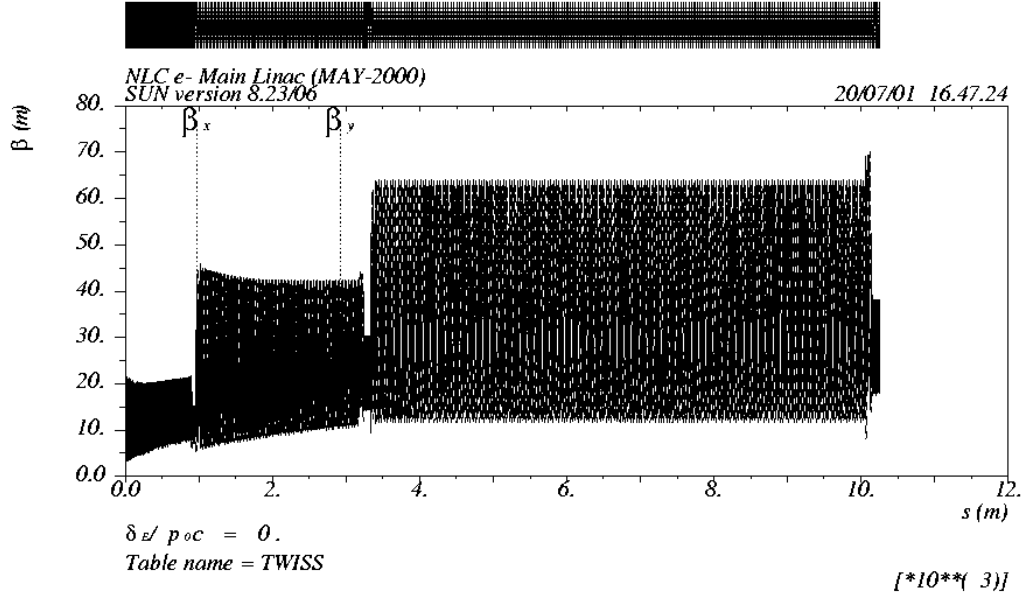}
\vspace{-0mm}
\caption{NLC main linac (version May 2000).}
\label{opt-nomin}
\vspace{4mm}
\end{figure}

NLC linac optics (version May 2000) and $\beta$ functions for a 500~GeV beam
are shown in Fig.~\ref{opt-nomin}.  The linac has three sectors consisting of
FODO cells with accelerator rf-structures placed between quadrupoles.
Compensation of wakefield effects requires stronger focusing at the low beam
energy.  This is achieved using short FODO cells in the first sector where
each half-cell contains three accelerator structures.  Cells in the second and
third sectors are longer and contain six and nine structures per half-cell,
respectively.  Phase advance per cell along the first and second sectors is
gradually decreased to take into account reduced wakefield effects with
energy.

\begin{figure}[!t]
\centering
\includegraphics*[width=125mm]{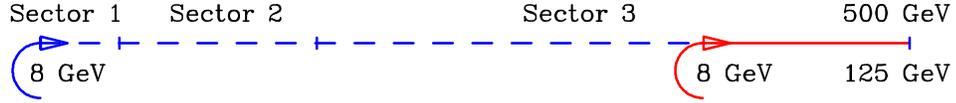}
\vspace{0mm}
\caption{Schematic of the nominal (dash+solid) and low energy linac (solid).}
\label{scheme}
\vspace{-0mm}
\end{figure}

Let's consider a scenario where the NLC is initially built for a lower beam
energy than in the nominal design.  To reduce the initial machine cost, only
the last part of the nominal linac may be installed for the low energy
operation.  We assume that injection can be provided at the beginning of the
shorter linac.  A schematic of the nominal (500~GeV) and low energy (125~GeV)
linacs is shown in Fig.~\ref{scheme}.  In this scenario, the low energy linac
would occupy part of the third sector of the full linac where the nominal
cells are long.  But for a wakefield compensation the low energy cells have
to be about three times shorter, as in the first sector of the linac.  It is
also desirable to keep the same lattice for the low and nominal energy
operations to reduce the cost of upgrade.

A compromise between the low and full energy optics can be achieved by a
simple modification of the nominal cell in the third sector.  The original
long cell which is described schematically as
\begin{eqnarray} 
\mathrm {F - 3rf - 3rf - 3rf - D - 3rf - 3rf - 3rf},
\label{cell-nomin} 
\end{eqnarray}
where F and D are focusing and defocusing quadrupoles, can be modified by
placing a quadrupole after every three accelerator structure module (3rf) to
generate either three identical short cells for a low energy linac:
\begin{eqnarray} 
\mathrm {F - 3rf - D - 3rf - F - 3rf - D - 3rf - F - 3rf - D - 3rf,} 
\label{cell-3short} 
\end{eqnarray} 
or one long cell for the nominal energy beam:
\begin{eqnarray} 
\mathrm {F - 3rf - F - 3rf - F - 3rf - D - 3rf - D - 3rf - D - 3rf.} 
\label{cell-1long} 
\end{eqnarray} 
Although the same physical quadrupoles are considered for both regimes, their
strengths will be different at low and high energy.  Cells in the second
sector may be modified similarly, if needed.

Modifications~(\ref{cell-3short})-(\ref{cell-1long}) add flexibility to the
linac optics, but require many more quadrupoles.  Based on the original
design, it is clear that the modified short cell~(\ref{cell-3short}) is
acceptable for the low energy beam since it's practically identical to the
nominal cell in the first sector.  The modified long cell~(\ref{cell-1long})
for the full energy beam is similar to the nominal long
cell~(\ref{cell-nomin}) except that F and D quadrupoles are split in three
pieces.  The quadrupole splitting increases the combined quadrupole strength
$KL$ and slightly reduces the maximum $\beta$ functions.

\begin{figure}[t]
\centering
\includegraphics*[width=80mm, angle=-90]{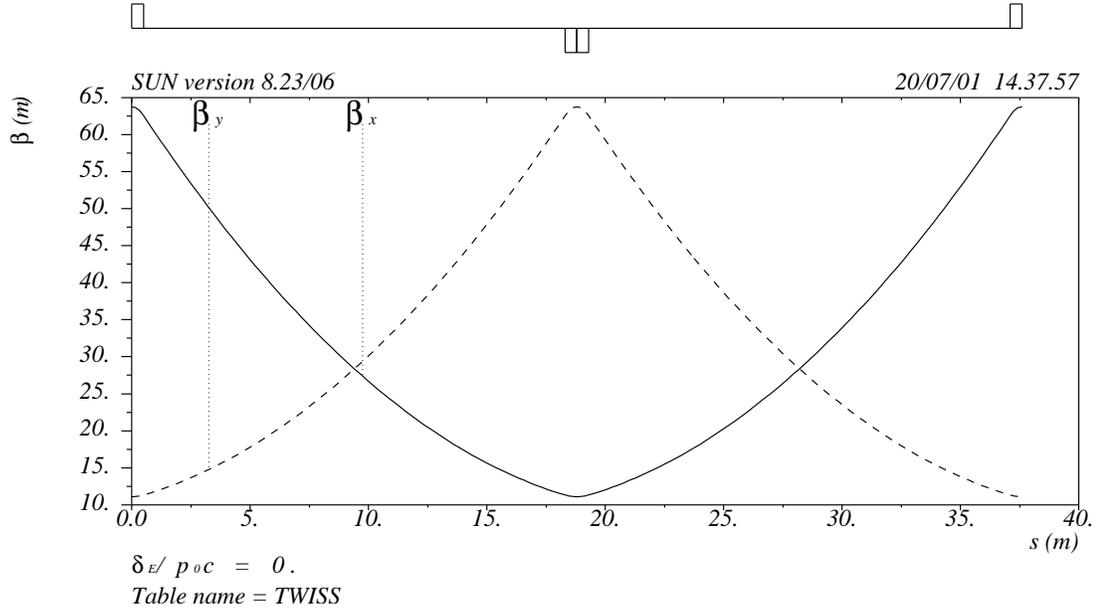}
\vspace{-0mm}
\caption{90 degree long FODO cell.}
\label{fd}
\vspace{4mm}
\end{figure}

\begin{figure}[!t]
\centering
\includegraphics*[width=80mm, angle=-90]{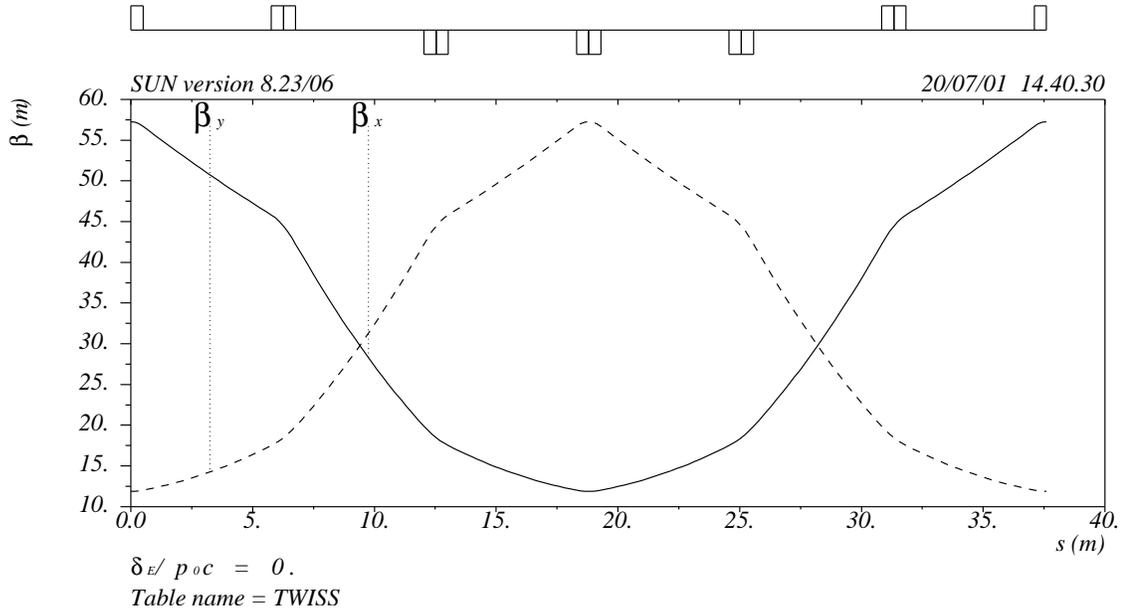}
\vspace{0mm}
\caption{90 degree FOFOFODODODO cell.}
\label{fffddd}
\vspace{-0mm}
\end{figure}

An example of the nominal and modified long cells in the third sector is shown
in Fig.~\ref{fd} and~\ref{fffddd}, where the cell phase advance of 90 degrees
is used.  The maximum $\beta$ functions are about 10\% lower in the modified
cell~(\ref{cell-1long}) in Fig.~\ref{fffddd}, but the combined quadrupole
strength $KL$ is up $\sim$60\% compared to the nominal cell.  The individual
quadrupoles in~(\ref{cell-1long}) are a factor of two weaker and therefore may
be as twice as short (for the same $K$) compared to the nominal quadrupoles.
Linear chromaticity $\frac{1}{4\pi}\sum\beta KL$ in the nominal and modified
cells is about the same:  -0.318 and -0.310, respectively.

Using design lattice for a 500~GeV beam (version May 2000), we estimated linac
length when quadrupoles are placed after every three structures (3rf).  In
this case, each 12.8 inches gap between 3rf modules in the nominal design is
replaced by a quadrupole with 6.275 inches space on both sides.  If the
original quadrupoles are used, then the modified second and third sectors
would be 77.7~m and 699.5~m longer, respectively, and the full linac would be
longer by 777.2~m.  However, if shorter quadrupoles are used in the modified
lattice, the length increase could be reduced by a half.

\begin{figure}[t]
\centering
\includegraphics*[width=80mm, angle=-90]{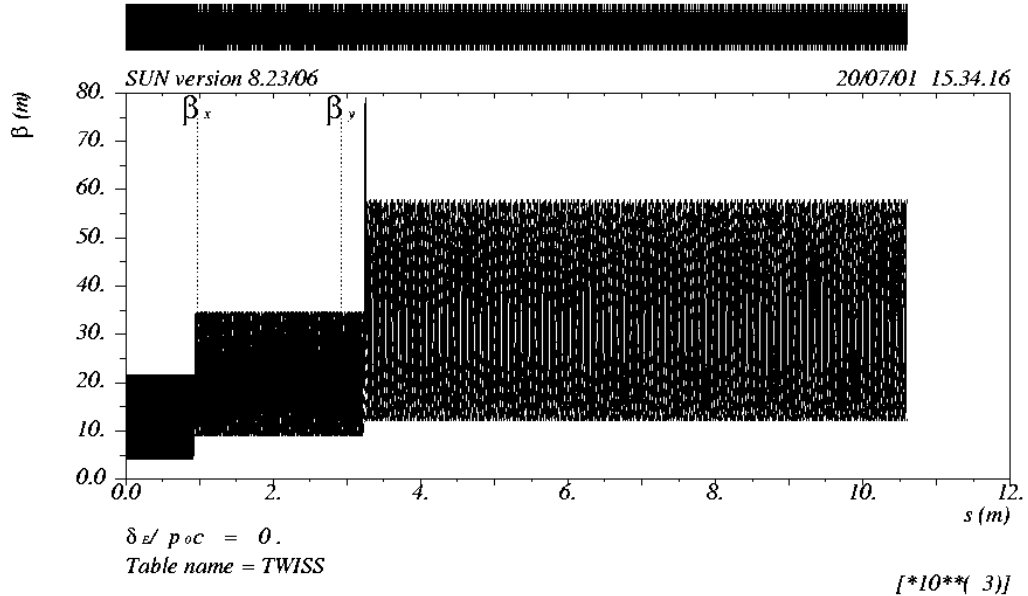}
\vspace{-0mm}
\caption{Modified 90 degree linac with a quadrupole after every 
three rf-structures.}
\label{opt-modif}
\vspace{-0mm}
\end{figure}

To generate a 500~GeV linac with a quadrupole after every three structures, we
used a specially written code:  ``Linac\_05''~\cite{linac5}.  It automatically
generates matched lattice for a three sector linac with F-3rf-D-3rf cells in
the first sector, F-3rf-F-3rf-D-3rf-D-3rf cells in the second sector, and
F-3rf-F-3rf-F-3rf-D-3rf-D-3rf-D-3rf cells in the third sector, where realistic
spaces between magnets and accelerator structures are used.  For an estimate,
we used phase advance of 90 degrees per cell in the modified lattice.  Short
12.75 inches quadrupoles were used throughout the modified linac to minimize
the total length.  Some other linac parameters were updated as well compared
to the original design.  The resultant $\beta$ functions are shown in
Fig.~\ref{opt-modif}.

Linac parameters for the original and modified lattices are listed in Table~1.
The nominal linac in Table~1 also includes a 114~m post-linac diagnostic
section while the modified linac does not.  The main disadvantage of the
modified lattice is that it requires a larger number of individual
quadrupoles.  However, Table~1 represents a conservative approach where both
sectors 2 and 3 are modified.  In principle, the nominal optics could be kept
unchanged in the second sector and beginning of the third sector which would
reduce the number of additional quadrupoles.

\begin{table}[!t]
\vspace{-0mm}
\begin{center}
\caption{Linac parameters for the nominal and modified lattices for 
500~GeV beam.}
\medskip
\begin{tabular}{|l|ll|}
\hline
& Nominal & Modified \\
\hline
Initial\,/\,final energy, GeV & 8\,/\,500 & 8\,/\,500 \\
Total length, m & 10257.924 & 10591.537 \\
Total accelerator structures & 4752 & 4977 \\
Total length of accel. structures, m & 8565.480 & 8975.522 \\
Length of one accel. structure, m & 1.8025 & 1.8034 \\
Accelerator gradient, MV/m & 57.440 & 54.816 \\
Total quadrupoles & 724 & 1660 \\
Total quadrupole length, m & 510.029 & 537.267 \\
Total quadrupole strength, kG & 420062 & 630325 \\
Maximum quadrupole gradient, kG/m & 1373.8 & 2021.0 \\
Total $x/y$ phase advance, [2$\pi$] & 79.68\,/\,84.12 & 88.45\,/\,88.87 \\
\hline
\end{tabular}
\end{center}
\vspace{0mm}
\end{table}

\section{Tolerances at 500~GeV/beam}

To compare alignment tolerances in the nominal and modified lattices for a
500~GeV beam, we computed effects of ground motion on beam emittance and beam
jitter using the ``Tolerance'' code~\cite{toler}.  This code analytically
calculates tolerances on quadrupole and rf-structure rms displacements for 6\%
emittance dilution and beam jitter amplitude of 25\% of beam size.  An
approximation is used where the entire space between quadrupoles is filled
with accelerator structures.  This way, the total length of structures is
slightly increased and the accelerator gradient is correspondingly reduced to
maintain the same energy gain.  The following beam parameters were used in the
tolerance calculation:
\begin{itemize} 
\vspace{-2mm} 
\item $3\!\times\!10^{-6}$ and $3\!\times\!10^{-8}$~m for $x$ and $y$ 
normalized emittance, respectively 
\vspace{-2mm} 
\item $1.1\!\times\!10^{10}$ particles per bunch 
\vspace{-2mm} 
\item 150 $\mu$m for rms bunch length 
\vspace{-2mm}
\item $1.29\!\times\!10^{20}$ V/C/m$^3$ for transverse wakefield slope in the
structures.  
\vspace{-2mm} 
\end{itemize}

The computed tolerances are shown in Table~2.  One can see that difference
between the nominal and modified linac tolerances are small.  Therefore, the
modified lattice is acceptable for running at the full beam energy.  This
lattice has more optical flexibility than the nominal lattice, but at the
expense of using many more quadrupoles.

\begin{table}[htb]
\vspace{-0mm}
\begin{center}
\caption{Ground motion tolerances for the nominal and modified 
linacs for 500~GeV beam.}
\medskip
\begin{tabular}{|lr|ll|}
\hline
& & Nominal & Modified \\
\hline
Quadrupole rms misalignment        & $x$ & 80.79 & 78.22 \\
for 6\% emittance dilution, $\mu$m & $y$ &  8.11 &  7.79 \\
\hline
Accel. structure rms misalignment  & $x$ & 14.90 & 15.38 \\
for 6\% emittance dilution, $\mu$m & $y$ &  1.52 &  1.54 \\
\hline
Quadrupole jitter for beam jitter  & $x$ & 75.01 & 83.40 \\
amplitude of 25\% of beam size, nm & $y$ &  7.45 &  8.35 \\
\hline
\end{tabular}
\end{center}
\vspace{0mm}
\end{table}

\section{Summary}

We described a modification of the NLC linac lattice suitable for initial low
energy beam running as well as an upgrade to the nominal energy.  In this
lattice, the long cells in the last segment of the linac are replaced by
shorter cells with quadrupoles placed after every three rf-structures.  The
modification is optimal for the low energy beam running and provides
acceptable ground motion tolerances at the nominal energy.  The main
disadvantages of the modified optics are a larger number of quadrupoles needed
for a full energy beam and up to 4\% longer linac.

\end{document}